# Series studies of the Potts model. III: The 3-state model on the simple cubic lattice (draft 24/12/93)

A J Guttmann† and I G Enting‡

†Department of Mathematics, The University of Melbourne, Parkville, Vic. Australia 3052.

‡CSIRO, Division of Atmospheric Research, Private Bag 1, Mordialloc, Vic. Australia 3195.

Short title: *Simple cubic three-state Potts series*

**Abstract.** The finite lattice method of series expansion has been used to extend low-temperature series for the partition function, order parameter and susceptibility of the 3-state Potts model on the simple cubic lattice to order $z^{43}$ and the high-temperature expansion of the partition function to order $v^{21}$. We use the numerical data to show that the transition is first-order, and estimate the latent heat, the discontinuity in the magnetisation, and a number of other critical parameters.

## 1. Introduction

This is the third in a series of papers in which we study the critical behaviour of the $q$-state Potts model in both two and three dimensions using series expansions derived from the finite lattice method. The first paper (Guttmann and Enting, 1993a), denoted I hereafter, gave the general expressions used to derive high- and low-temperature expansions for the $q$-state Potts model. In I, series expansions for the $q = 2$ (Ising) case on the simple cubic lattice were analysed. The second paper (Briggs et al 1994) denoted II hereafter, presented and analysed series for the bulk thermodynamic properties for Potts models on the square lattice for integer $q$ ranging from 2 to 10. These were used to develop and test series analysis techniques that could distinguish between first-order and continuous transitions. The present paper considers the three-state model on the simple cubic lattice.

After the initial paper by Potts (1952), the model attracted little attention for almost two decades. During the 1970's there was greatly renewed interest in the model, with new exact results, series studies and renormalisation group calculations and applications to phase transitions in surface films. A particular concern at that time was the failure of renormalisation group calculations to reproduce the exact results for the order of the transition in two dimensions. A review by Wu (1982) described much of the work on the Potts model.

Of even greater interest is the behaviour of the three-dimensional Potts model. As noted above, for the $q = 2$ (Ising) case, the low-temperature series and some high-temperature series have recently been extended in I. For $q = 3$, the three-dimensional Potts model is of particular interest as it is in the same universality class as the Z(3) clock model. This in turn is the centre of SU(3), so it is believed that the effective theory for Polyakov loops in finite-temperature $d = 4$ SU(3) lattice gauge theory should be in the same universality class as the three-dimensional three-state Potts model. The Polyakov loops are the order parameter for the deconfinement transition that is thought to take place in QCD as the temperature is raised (hadronic quarks going over to a plasma of free quarks and gluons). The connection then is that if the $d = 3$, $q = 3$ Potts model has a first-order transition then the above SU(3) transition must also be first-order. On the other hand if the Potts model has a continuous transition then the SU(3) transition could be either first-order or continuous. A number of earlier studies (Knak Jensen *et al* , 1979, Nienhuis *et al* 1981, Blöte and Swendsen, 1979, Kim and Joseph, 1975, Kogut and Sinclair, 1981, Enting, 1974, Herrman, 1979) claimed to see evidence of a first-order transition for the three-dimensional 3-state model, but these were based on Monte Carlo analyses on small lattices, or series work on short series. In 1988, results of large scale QCD calculations by Bacilieri *et al* (1988) and by Brown and Christ (1988) gave conflicting results, with the former claiming to find evidence of a second-order phase transition, while the latter claimed to find evidence of a first-order transition. Further Monte Carlo studies by Fukugita *et al* (1990) supported the first-order results, but Cabasino *et al* (1989) argued that the evidence was equally good for a first-order or continuous transition. Brown (1989) argued for a first-order transition, a result supported by Gavai and Karsch (1992), who considered the effect of additional next-nearest neighbour couplings. Further, Alves *et al* 1991 also used high precision Monte Carlo methods to conclude that the transition was first-order. Very recently, Bhanot *et al* (1993) used a method similar to ours to extend the Potts model series, and analysed the series to provide additional evidence for a first-order transition. After this work was completed, Vohwinkel (1993) showed how the shadow-graph method of Sykes (1965) could be used to extend the series even further than we have. However Vohwinkel has only presented the magnetisation series, and no analysis.

In order to distinguish between first-order and continuous transitions we have analysed the series using differential approximants (Guttmann, 1989) to integrate the series. The studies of the square lattice system in II gave a procedure for determining the order of the transition in this way. We have used these ideas to locate the critical temperature, and to identify the nature of the transition.

The layout of the remainder of the paper is as follows: In the next section we briefly describe the finite lattice method and the nature of the results we have obtained therefrom. In section 3 we analyse the data. In section 4 we present a discussion of the results.

## 2. Series expansions from the finite lattice method

The definitions and notation follow the usage of I (and II). The standard $q$-state Potts model is defined on a lattice with each site having a 'spin' variable that takes on $q$ possible values (denoted '0' to $q-1$). An energy $\Delta E$ is associated with each pair of interacting sites that are in different spin states, and an energy of 0 applies to pairs of interacting sites in the same state. We consider the simple cubic lattice, with each site interacting only with its 6 nearest neighbours. Each site not in state '0' has an additional field energy $H$.

The thermodynamic quantities can be derived from the partition function. We choose the normalisation such that the state with all sites in state '0' has zero energy. In this normalisation, the partition function is commonly denoted $\Lambda$.

We work in terms of the expansion variables $z = \exp(-\Delta E/kT)$, $\mu = \exp(-H/kT)$ and the high-temperature variable $v = (1-z)/(1+(q-1)z)$

For the simple cubic lattice, the high-temperature expansion for the partition function takes the form
$$\Lambda = q^{-2}(1+(q-1)z)^3\,\Phi(v), \tag{1}$$
with
$$\Phi(v) = \sum_n a_n v^n = 1 + 3(q-1)v^4 + \dots.$$

For the low-temperature expansion, we use a modified field variable $x = 1 - \mu$ and truncate at order $x^2$ so that the partition function is expressed as
$$\Lambda = \Lambda_0 + x\Lambda_1 + x^2\Lambda_2 + \dots. \tag{2}$$

The expansion of the zero-field partition function is written as
$$\Lambda_0 = \sum_n \lambda_n z^n. \tag{3}$$

The internal energy is given by
$$U = z\frac{d\Lambda_0}{dz}/\Lambda_0. \tag{4}$$

the order parameter by
$$M = 1 + \frac{q}{q-1}\frac{\Lambda_1}{\Lambda_0} = \sum_n m_n z^n, \tag{5}$$

and the susceptibility by
$$\chi = 2\frac{\Lambda_2}{\Lambda_0} - \frac{\Lambda_1}{\Lambda_0} - \left(\frac{\Lambda_1}{\Lambda_0}\right)^2 = \sum c_n z^n. \tag{6}$$

Note that for $q \geq 3$ an additional 'transverse' susceptibility can be defined (Straley and Fisher, 1973).

Previously, series expansions for the Potts model on the simple cubic lattice had been obtained by Straley (1974) ( for low-temperature $\ln Z$, $M$ and $\chi$ to $z^{24}$ and high-temperature series quoted to $v^{10}$) and by Miyashita et al (1979) ($\ln Z$ with full field-dependence to order $z^{33}$ and full temperature dependence to $\mu^{11}$) In 1993 Bhanot et al extended the free-energy and magnetisation series to 39 terms, and the susceptibility series to 35 terms. Vohwinkel (1993) has given the magnetisation series to 56 terms, and has (unpublished) similar length series for other thermodynamic properties.

The description of the finite lattice calculations in I was couched in terms of the general $q$-state Potts model and it was that formalism that was used in the present study. As noted in I (see also II, equation 17a and b), the amount of memory required increases with $q$. The present calculations used cuboids with cross-sections of up to $3 \times 4$ sites, giving the low-temperature series correct to $z^{41}$. As described in I, comparing the $3 \times 4$ approximation for $q = 2$ to higher-order $q = 2$ calculations allows us to determine the correction required to the $3 \times 4$ approximation for all other $q$ values. This has enabled us to extend the low-temperature series to $z^{43}$.

For the high-temperature series, we ran on a $4 \times 4$ lattice, which required storage of 900 MB. This gave series correct to $v^{21}$. All programs were run on an IBM 3090/400J with 1/2 GB of memory and 2 GB of backing storage. One run was also performed on a Cray EL. The runs for the high-temperature series took about 50 hours, the low-temperature $3 \times 4$ lattice runs substantially less. To extend the low-temperature series would currently take 4 GB of memory. However, for low-temperature series, we believe that the shadow-lattice method is computationally superior, so there seems little incentive to develop this method further for low-temperature Potts series. We discuss this point further in Guttmann and Enting (1993b).

The coefficients for $\lambda_n$, $m_n$, $c_n$ and $a_n$ for $q = 3$ are listed in Table 1. These series disagree in several places with previously published series. The low-temperature series disagree at $z^{24}$ with those published by Straley (1974). This error had previously been detected using series obtained by the method of partial generating functions (Enting, unpublished) and in fact the error was a typographical mistake; the analysis by Straley used the correct series (Straley, personal communication). Apparently the term quoted in the appendix to Straley (1974) as $486(y_1^6 + y_2^6)x^{24}$ should have been $496(y_1^6 + y_2^6)x^{24}$.

The low-temperature series also disagree with those of Miyashita *et al* (1979) at order $z^{30}$. We have not been able to obtain the full field-dependent corrections to their series, but note that their coefficient for $z^{30}\mu^5$ should be $(q-1)^5$ times the corresponding Ising coefficient (see Sykes *et al*, 1965), i.e. it should be $44998\frac{2}{5}$ and not 45008 as published. As our coefficients agree with those of Bhanot *et al* and Vohwinkel, we are confident that they are correct.

The most serious discrepancies are between our new series and the high-temperature series published by Straley (1974), disagreeing at order $v^8$ and $v^{10}$. In view of the gross disagreements at quite low order, we present, in an appendix to this paper, an independent re-calculation of these series for general $q$, using a conventional weak-graph expansion. This expansion confirms our finite lattice calculations and also reproduces the series for the mean number of clusters in bond percolation. The expansion also agrees with Ising model series from I, but this is a weaker test of either our weak-graph expansion or the finite lattice method expansion because many graph types have zero contribution for $q = 2$ and, in particular, only even powers of $v$ occur. Nevertheless, this appendix provides useful series for general q, and is likely to be useful for other workers as a check on any long series that may subsequently be obtained.

## 3. Analysis of series

The series generated as described above were all analysed by the method of differential approximants (DA) (Guttmann, 1989, page 83ff). This method generalizes Padé approximants by fitting an ordinary differential equation of the form

$$\sum_{i=0}^{m} Q_i(x)D^i f(x) = P(x)$$

(with $d^i$ denoting $(\frac{\mathrm{d}}{\mathrm{d}x})^i$) to the available series terms. Here $Q_k(x) = \sum_{i=0}^{m_k} q_{ki}x^i$ and $P(x) = \sum_{i=0}^{m_0} p_i x^i$ are polynomials. We chose $q_{m0} = 1$, so that the origin is not a regular singular point. This allows integration of the differential equation starting at $x = 0$. (This then corresponds to logarithmic derivative Padé approximants when $m = 1$). For magnetisation series, homogeneous DAs ($P \equiv 0$) are often most useful. Generally, the degrees of $Q_k$ and $P$ are chosen to use all (or most) of the available series terms. In principle any order of differential equations can be used, but first-order ($m = 1$) was mostly used in the current work. Finding the coefficients of $Q_k$ and $P$ reduces to the solution of a system of linear equations, but this system is often ill-conditioned, so that care must be taken in its solution.

These differential equations were then integrated numerically to obtain estimates of the desired physical quantities. In all cases a number (up to 10) of DAs were integerated and the results averaged to obtain the means and standard deviations shown in the tables and graphs below. All calculations were performed in quadruple precision (approximately 34 decimal places), so that all series terms could be represented without loss of precision.

We performed the numerical integration with an extrapolation method of the Bulirsch-Stoer type, as described by Hairer (1987, Section II.9).

The analysis in II involved applying these methods to square-lattice series for comparison with the exact results of Baxter (1973, 1982) to asses the extent to which the methods could distinguish between first-order and continuous transitions. We showed that we could clearly distinguish the order of the transition in all known cases.

Simple sequence transformations were used to generate the most appropriate series, and hence DA, for numerical integration. In general, if a quantity is believed to behave like $T^k$ at the origin of integration, it is often useful to transform the series $\sum_{i=0}^{N} a_i T^i$ say, to $\sum_{i=0}^{N} a_i T^i / T^k$ so that the transformed function approaches a constant at the origin. Thus, for example, we worked with the series for $\chi/z^6$, rather than $\chi$ itself. Similarly, instead of the magnetisation $M = 1 - 3z^6 - 18z^{10} - 18z^{11}..$, we worked with $M - 1 + 3z^6$. We now discuss our numerical results in greater detail.

*3.1. Internal energy*

We integrated the internal energy series $U(z)$ from $T = 0$ and $T = \infty$ until they crossed at $T_c$. The results are shown in Fig. 1, with the intersection region shown as an inset. It is perfectly clear already, from this graph alone, that the transition is first-order. This follows from the fact that the gradient is clearly discontinuous. The intersection of the two curves gives the critical temperature. From a range of several approximants, we find the intersection at $kT/J = 1.8168 \pm 0.0012$, which compares with Monte Carlo estimates of $1.8166 \pm 0.0002$ (Yamagata 1993), $1.816454 \pm 0.000032$

(Alves *et al* 1991), $1.8164 \pm 0.0001$ (Fukugita *et al* 1990), $1.8161 \pm 0.0001$ (Gavai *et al* 1989) and $1.81624 \pm 0.00006$ (Wilson and Vause, 1987).

It is clear that these estimates do not all agree within the stated precision. However it is also clear that the Monte Carlo estimates are all slightly lower than our estimate. The average of the Monte Carlo estimates is taken to be 1.8163, and all subsequent analysis will be performed using both our central estimate of $T_c$ and the Monte Carlo average. Note too that the MC estimate is made *under the assumption of a first-order transition*, while our series analysis makes no such assumption.

By differentiation, we can readily construct series for the high- and low-temperature internal energy. Integration of these gives a latent heat of $0.264 \pm 0.011$, where the error is one standard deviation in the average of the approximants. This error swamps the error induced by the uncertainty of the critical temperature. Our normalisation of the Hamiltonian has, as a consequence, that the internal energy varies between 0 and 2. Other workers use a different normalization, in which the internal energy varies between 0 and 1. Therefore their latent heat estimates must be doubled to be compared with that given here. The specific heat at $T_c^-$ is found to be $30 \pm 4$, while $T_c^+$ was considerably lower, at $11.1 \pm 0.6$. For a first order transition, the specific heat is undefined at $T_c$, though the left- and right-hand limits are of course defined. By use of Monte Carlo methods, Gavai *et al* (1989) find a latent heat of $0.16 \pm 0.008$, Alves *et al* find $0.1606 \pm 0.0006$ while Gavai and Karsch (1992) find a latent heat of $0.160 \pm 0.07$. Our result is some 60% higher than these estimates. More precisely, we find $E(T_c^-) = 1.151 \pm 0.009$ and $E(T_c^+) = 1.414 \pm 0.004$. We have no explanation for the difference between our results and the Monte Carlo results. However, we note that our methods did give the correct latent heat in the two-dimensional case. That is, all the exact values lay within the range defined by the set of approximants with extreme outliers removed.

*3.2. Magnetisation*

As $T_c^-$ is approached, the gradient decreases rapidly. Consequently, the magnetisation gap depends critically on the estimate of the critical temperature. Using the MC value of $T_c$, we find $\Delta M = 0.498$, while using our estimate of $T_c$, we find $\Delta M = 0.505$. Another factor is the length of the series. Extending the series by 13 terms, as has been done by Vohwinkel, gives $\Delta M = 0.463$ at the Monte Carlo value of $T_c$, a drop of 7%. A similar decrease is found using the extended series at the other value of $T_c$ used.

Monte Carlo estimates of this quantity have been obtained by Gavai *et al* (1989), who find $\Delta M = 0.395 \pm 0.005$, only a little lower than our estimate.

In Table 4 we show the location and residues of Dlog Padé approximants to the magnetization series. They indicate a critical temperature of $e^{J/kT_c} \approx 0.5785$, some 0.3% above the value we believe to be correct. A "pseudo magnetization exponent" around 0.2 is also suggested. As shown in II, this is entirely consistent with the existence of a first-order transition.

*3.3. Susceptibility*

As for the specific heat, one expects the susceptibility at a first-order transition to be undefined, while having well-defined left- and right-hand limits. As we only have low-temperature susceptibility series, we can only find one limit. We find a large but finite value of the susceptibility at $T_c^-$, notably $13 \pm 3$. on the low-temperature side.

**4. Discussion of results**

Hamer *et al* (1992) have studied the quantum Hamiltonian version of the 3-state Potts model in $(2+1)$ dimensions. They find $\Delta M = 0.42 \pm 0.02$, and a latent heat jump of 0.21 or 0.24, depending on the lattice. Thus we see that the three distinct methods of study, series analysis and Monte Carlo of the Potts model, and series analysis of the quantum analogue of the Potts model, give consistent results. That is, they all find a fairly weak first-order transition, with a small latent heat, but quite a large jump in the magnetisation.

We have shown how the finite-lattice method can provide competitive series in three-dimensions, though the computational complexity ensures that state-of-the-art direct methods, well-programmed, will eventually be superior. And as the dimensionality increases, the method becomes steadily worse (Guttmann and Enting, 1993b).

We have also shown that series methods can provide not only qualitative information about the nature of a phase transition, but also quantitative estimates of critical parameters for first-order, as well as second-order phase transitions.

Our methods have still provided by far the longest extant high-temperature series, and this is needed to determine the nature of the phase transition, as well as to identify its location.

**Appendix 1. Comparison with weak-graph expansions**

As noted in section 2, our high-temperature series shows significant differences from that given by Straley (1974). Since the program used to produce our series (and those of I) is our first use of a site representation for calculating high-temperature series, these differences caused us some concern. In order to check our program we recalculated the free energy using a conventional weak-graph expansion.

The graphs that are required are those with no vertices of order 1 and in which the removal of one line does not increase the number of components. Each graph requires a $q$-dependent weighting factor. This can be assigned by assigning each directed bond an integer flow in the range 1 to $q-1$. At each vertex the sum of inwards flows minus outwards flows must be 0 modulo $q$ (see for example, Straley, 1974). The weight is the number of ways such flows can be assigned. An alternative algebraic formulation is given by Domb (1974). For planar graphs, the weighting is equivalent to $q^{-1}$ times the number of $q$-colourings of the graph and its exterior (see for example Wu, 1982).

Table 3 gives details of the expansion. To save space, the lowest order terms are not shown in the Table. They are, at fourth order, 3 polygons, at sixth order, 22 polygons and at seventh order, 18 theta graphs. These then give the following low-order terms, with subsequent terms being given in Table 3:

$\frac{d}{dq} \ln \lambda(q=1) = 3x^4 + 22x^6 - 18x^7 + \ldots$,

$\ln \Lambda(2) = 3x^4 + 22x^6 + 0x^7 + \ldots$,

$\ln \Lambda(3) = 6x^4 + 44x^6 + 36x^7 + \ldots$,

$\ln \Lambda(4) = 9x^4 66x^6 + 108x^7 + \ldots$,

$\ln \Lambda(5) = 12x^4 + 88x^6 + 216x^7 \ldots$.

The results of the finite lattice calculations agree with the weak graph expansion for $\ln \Lambda$. Further the $q=2$ series (see I) reproduces the known Ising model series.

An interesting additional check is that our general-$q$ expansion enables us to calculate $K(p)$ the mean number of clusters in the low-density regime of the bond percolation problem. Bond percolation can be regarded as the $q \to 1$ limit of the Potts model (Fortuin and Kasteleyn, 1972, or, more accesibly, Wu, 1978). Sykes (1986) give series for site clusters which agree with our series, after taking into account the cluster of 1 site and zero bonds that is included in the Potts model limit, but excluded from conventional enumerations of bond percolation.

The agreement between the series of Sykes and the results derived from the weak graph expansion of the Potts model give a useful test of our tabulation because the original derivation of the percolation series used a quite different type of expansion.

## Acknowledgments

We would like to thank Dr. Alan Sokal for urging us to do this calculation and for explaining to us the significance in the context of lattice gauge theories as discussed in the introduction. We would also like to thank Dr. Martin Sykes for the provision of the weak lattice constants used to construct Table 3. The assistance of Dr. Robert Bursill and Mr. Keith Briggs in the numerical integration is gratefully acknowledged. Financial support from the Australian Research Council is also gratefully acknowledged. The support of the Australian Computing and Communications Institute, who provided the computational facilities in the form of an IBM 3090/400J is greatly appreciated. In particular Mr. Glenn Wightwick and Mr. Jan Jager helped to run our very large jobs.

# References

Alves N A, Berg B A and Villanova R 1991 *Phys. Rev. B* **43** 7

Bacielieri P et al.1988 *Phys. Rev. Lett.* **61** 1545

Baxter R J 1973 *J. Phys. C* **6** L445

Baxter R J 1982 *J. Phys. A: Math. Gen.* **15** 3329

Bhanot G, Creutz M, Glässner U, Horvath I, Lacki J, Schilling K and Weckel J 1993 *Phys. Rev. B* **48** 6183

Briggs K, Enting I G and Guttmann AJ 1994 Series studies of the Potts model II. Bulk series for the square lattice. *J. Phys. A: Math. Gen.* **27** (to appear)

Brown F R 1989 *Phys. Lett. B* **224** 412

Brown R R and Christ N H 1988 *Phys. Rev. Lett.* **61** 2058

Cabasino S et al 1990 *Nucl. Phys. B, Proc. Supp. Section* **17** 218-222

Domb C 1974 *J. Phys. A: Math. Gen.* **7** 1335–48

Enting I G 1974 *J. Phys. A: Math. Gen.* **7** 1617

Fortuin C M and Kasteleyn P W 1972 *Physica* **57** 536

Fukugita M, Mino H, Okawa M and Ukawa A 1990 *J. Stat. Phys.* **59** 1397

Gavai R V and Karsch F 1992 *Phys. Rev. B* **46** 2

Gavai R V, Karsch F and Petersson B 1989 *Nucl. Phys.* **B322** 738

Guttmann A J 1989 Asymptotic Analysis of Power-Series Expansions. In C. Domb and J. L. Lebowitz (eds.) *Phase Transitions and Critical Phenonema. Volume 13.* (Academic Press 1989).

Guttmann A J and Enting I G 1993a *J. Phys. A: Math. Gen.* **26** 807

Guttmann A J and Enting I G 1993b *Phys. Rev. Lett.* **70** 698

Hairer E et al 1987 *Solving Ordinary Differential Equations I, Nonstiff Problems.* Springer Series in Computational Mathematics 8.

Hamer C J, Oitmaa J and Weihong Z 1992 *J. Phys. A: Math. Gen.* **25** 1821–1833

Herrman H J 1979 *Z. Physik B* **35** 171

Kim D and Joseph R I 1975 *J. Phys. A: Math. Gen.* **8** 891

Knak Jensen S J and Mouritsen O G 1979 *Phys. Rev. Lett.* **43** 1736

Kogut J B and Sinclair D K 1981 *Phys. Letts.* **81A** 149

Miyashita S, Betts D D and Elliott C J 1979 *J. Phys. A: Math. Gen.* **12** 1605-22

Neinhuis B, Reidel E K and Schick M 1981 *Phys. Rev. B* **23** 6055

Potts R B 1952 *Proc. Camb. Phil. Soc.* **48** 106

Straley J P 1974 *J. Phys. A: Math. Gen.* **7** 2173–80

Straley J P and Fisher M E 1973 *J. Phys. A: Math. Gen.* **6** 1310–26

Sykes M F, Essam J W and Gaunt D S 1965 *J. Math. Phys.* **6** 283–298

Sykes M F, Essam J W, Heap B R and Wiley B J 1966 *J. Math. Phys.* **7** 1557

Sykes M F 1986 *J. Phys. A: Math. Gen.* **19** 2425

Vohwinkel C 1993 *Phys. Lett. B* **301** 208–212

Wilson W G and Vause C A 1987 *Phys. Rev. B* **36** 587

Wu F Y 1978 *J. Statist. Phys.* **18** 115–23

Wu F Y 1982 *Rev. Mod. Phys.* **54** 235–68

Yamagata A 1993 *J. Phys. A: Math. Gen.* **26** 2091

# Tables and table captions

| $n$ | $\lambda_n$ | $m_n$ | $c_n$ |
|---|---|---|---|
| 0  | 1          | 1           | 0           |
| 6  | 2          | -3          | 2           |
| 10 | 6          | -18         | 24          |
| 11 | 6          | -18         | 24          |
| 12 | -12        | 42          | -56         |
| 14 | 30         | -135        | 270         |
| 15 | 60         | -270        | 540         |
| 16 | -96        | 477         | -930        |
| 17 | -132       | 648         | -1296       |
| 18 | 346        | -1980       | 4768        |
| 19 | 498        | -2988       | 7968        |
| 20 | -636       | 4140        | -10560      |
| 21 | -2210      | 14052       | 36922       |
| 22 | 3000       | -21690      | 64812       |
| 23 | 7344       | -52920      | 163440      |
| 24 | -7110      | 55020       | -165464     |
| 25 | -25836     | 201852      | -659088     |
| 26 | 17802      | -162774     | 600024      |
| 27 | 107450     | -914538     | 3278256     |
| 28 | -59358     | 555750      | -1980408    |
| 29 | -353376    | 3229524     | -12285816   |
| 30 | 105944     | -1188327    | 5005014     |
| 31 | 1342914    | -13301370   | 55200864    |
| 32 | -77154     | 1402686     | -6062712    |
| 33 | -4995004   | 52334268    | -227203096  |
| 34 | -226914    | 95751       | 1954650     |
| 35 | 17383710   | -195398208  | 914339736   |
| 37 | -64127562  | 761838084   | -3742275288 |
| 38 | -32638848  | 359664885   | -1761828642 |
| 39 | 231546628  | -2910516786 | 15132717432 |
| 40 | 160963416  | -1946958399 | 10380877350 |
| 41 | -805061298 | 10681132140 | -58385376120 |
| 42 | -795051840 | 10207745148 | -56515869708 |
| 43 | 2914349712 | -40522674258 | 232316142012 |

**Table 1.** Coefficients in low-temperature expansions for $\Lambda_0$, $M$ and $\chi$, defined by equations (3), (5) and (6).

| $n$ | $a_n$ |
|---|---:|
| 0 | 1 |
| 4 | 6 |
| 6 | 44 |
| 7 | 36 |
| 8 | 402 |
| 9 | 688 |
| 10 | 4836 |
| 11 | 11364 |
| 12 | 69466 |
| 13 | 196374 |
| 14 | 1097436 |
| 15 | 3583084 |
| 16 | 18627090 |
| 17 | 67523316 |
| 18 | 335693618 |
| 19 | 1305112008 |
| 20 | 6332595828 |
| 21 | 25841846466 |

**Table 2.** Coefficients in the high-temperature expansion for $\Phi$, defined by equation (1).

| Graph | W | 8 | 9 | 10 | 11 | 12 | 13 | 14 |
|---|---|---|---|---|---|---|---|---|
| p | 1 | 207 | | 2412 | | 31754 | | 452640 |
| $\theta$ | Q | 24 | 344 | 528 | 5934 | 12120 | 104250 | 239610 |
| $\alpha$ | QT | | 8 | 24 | 228 | 996 | 5916 | 29448 |
| $\beta$ | $Q^2$ | | | 60 | 96 | 2556 | 6480 | 76752 |
| $\gamma$ | $Q^2$ | | | 84 | 288 | 2400 | 9552 | 52584 |
| $\delta$ | $Q^2 + S$ | | | | 12 | 279 | | 5388 |
| F | $W_F$ | | | | | 15 | 168 | 1320 |
| J | $Q(Q^2 + 1)$ | | | | | 48 | 192 | 1560 |
| B | $QT^3$ | | | | 12 | | 384 | 756 |
| 117 | $W_{117}$ | | | | | 1 | | 24 |
| G | $Q^2T$ | | | | | 96 | 336 | 4392 |
| 8=(p.p) | S | 30 | | 576 | | 9306 | | 152784 |
| $(\theta.p)$ | QS | | | | 456 | 648 | 14796 | 24900 |
| (8.p) | $S^2$ | | | | | 458 | | 14460 |
| $[\theta.p]$ | QS | | | | -828 | -1224 | -28560 | -49656 |
| [8.p] | $S^2$ | | | | | -1620 | | -54456 |
| [p,p] | S | $\frac{-99}{2}$ | | $-1020$ | | $-17510$ | | $-300108$ |
| [p,p,p] | $S^2$ | | | | | 1377 | | 48642 |
| $[\theta, \theta]$ | $Q^2S$ | | | | | | | -3384 |
| $(\theta.\theta)$ | $Q^2S$ | | | | | | | 1674 |
| $[\alpha, p]$ | QST | | | | | | | -408 |
| $(\alpha.p)$ | QST | | | | | | | 216 |
| $[\beta, p]$ | $Q^2S$ | | | | | | | -3450 |
| $(\beta.p)$ | $Q^2S$ | | | | | | | 1704 |
| $[\gamma, p]$ | $Q^2S$ | | | | | | | -4956 |
| $(\gamma.p)$ | $Q^2S$ | | | | | | | 2604 |
| $[\delta, p]$ | $Q^2S$ | | | | | | | -708 |
| $(\delta.p)$ | $Q^2S$ | | | | | | | 384 |
| 41 | $W_{41}$ | | | | | | | 12 |
| 69 | $W_{69}$ | | | | | | | 96 |
| 70 | $W_{70}$ | | | | | | | 72 |
| 78 | $W_{78}$ | | | | | | | 24 |
| 79 | $W_{79}$ | | | | | | | 12 |
| 93 | $W_{93}$ | | | | | | | 96 |
| 99 | $W_{99}$ | | | | | | | 24 |
| C | $Q^2T$ | | | | | | 72 | 1344 |
| D | $Q^3$ | | | | | | 150 | 300 |
| H | $Q^3$ | | | | | | 600 | 2112 |
| I | $Q^3$ | | | | | | | 264 |
| K | $Q^3$ | | | | | | 204 | 1248 |
| L | $Q^3$ | | | | | | 228 | 240 |
| M | $W_M$ | | | | | | 108 | 72 |
| N | $Q^3$ | | | | | | 144 | 1008 |
| O | $W_O$ | | | | | | | 144 |
| P | $W_M$ | | | | | | 120 | 144 |
| Q | $W_Q$ | | | | | | 3 | |

| Graph | $W$ | 8 | 9 | 10 | 11 | 12 | 13 | 14 |
|---|---|---|---|---|---|---|---|---|
| $\frac{d}{dq}\ln\Lambda(q=1)$ | | 183 | -328 | 2034 | -5142 | 26539 | -81183 | 381222 |
| $\ln\Lambda(2)$ | | $\frac{375}{2}$ | | 1980 | | 24044 | | 319170 |
| $\ln\Lambda(3)$ | | 384 | 688 | 4572 | 11184 | 66158 | 190662 | 1050924 |
| $\ln\Lambda(4)$ | | $\frac{1179}{2}$ | 2112 | 8856 | 34956 | 162624 | 693192 | 3153690 |
| $\ln\Lambda(5)$ | | 804 | 4320 | 15912 | 73224 | 358220 | 1573821 | 7815144 |

**Table 3.** Details of high-temperature expansion in terms of weak lattice constants. Body of table gives lattice constant for 8 to 14 bonds on SC lattice. (The contribution from lower-order graphs is given in the Appendix). Column 1 identifies graph type, in the notation of Sykes *et al* 1966. $(a.b)$ denotes a type $a$ and type $b$ graph with a common vertex. $[a,b]$ is used to denote types $a$ and $b$ as separate components. The '8' denotes $(p.p)$, the 'figure 8' type. Column 2 is the weight, with a common factor of $q-1$ removed. Here $S$ denotes $(q-1)$, $Q$ denotes $(q-2)$, $T$ denotes $(q-3)$, $W_F = Q(q^2 - 5q + 7)$ and $W_{117} = Q(q^3 - 9q^2 + 29q - 32), W_{41} = Q(q^3 - 6q^2 + 149 - 13), W_{69} = QT(q^2 - 4q + 5), W_{70} = Q^2(q^2 - 5q + 8), W_{78} = Q(q^3 - 7q^2 + 18q - 17), W_{79} = Q^2(q^2 - 5q + 8), W_{93} = (QT)^2, W_{99} = QT(q^2 - 5q + 7), W_M = Q(q^2 - 3q + 3), W_0 = (q^3 - 5q^2 + 10q - 7), W_Q = Q(q^2 - 2q + 2)$.

| $N$ | $[N-1/N]$ | | $[N/N]$ | | $[N+1/N]$ | |
|---|---|---|---|---|---|---|
| 16 | 0.57804 | (0.1990)* | 0.57822 | (0.2008)* | 0.57870 | (0.2054) |
| 17 | 0.57908 | (0.2090) | 0.57852 | (0.2037)* | 0.57855 | (0.2040) |
| 18 | 0.57856 | (0.2041) | 0.57838 | (0.2025)* | 0.57856 | (0.2041)* |
| 19 | 0.57857 | (0.2042)* | 0.57855 | (0.2040)* | 0.57934 | (0.2031)* |
| 20 | 0.57848 | (0.2033) | 0.57847 | (0.2031) | | |

**Table 4.** $q=3$, Dlog Padé approximants to the simple-cubic lattice spontaneous magnetisation, giving the location and residue of the "pseudo critical point".

**Figure captions**

**Figure 1.** Free-energy versus $Tk/J - T_c k/J$, where the Monte Carlo estimate of $T_c k/J = 1.8163$ has been used. The inset figure shows a blow-up of the intersection of the high- and low-temperature curves.

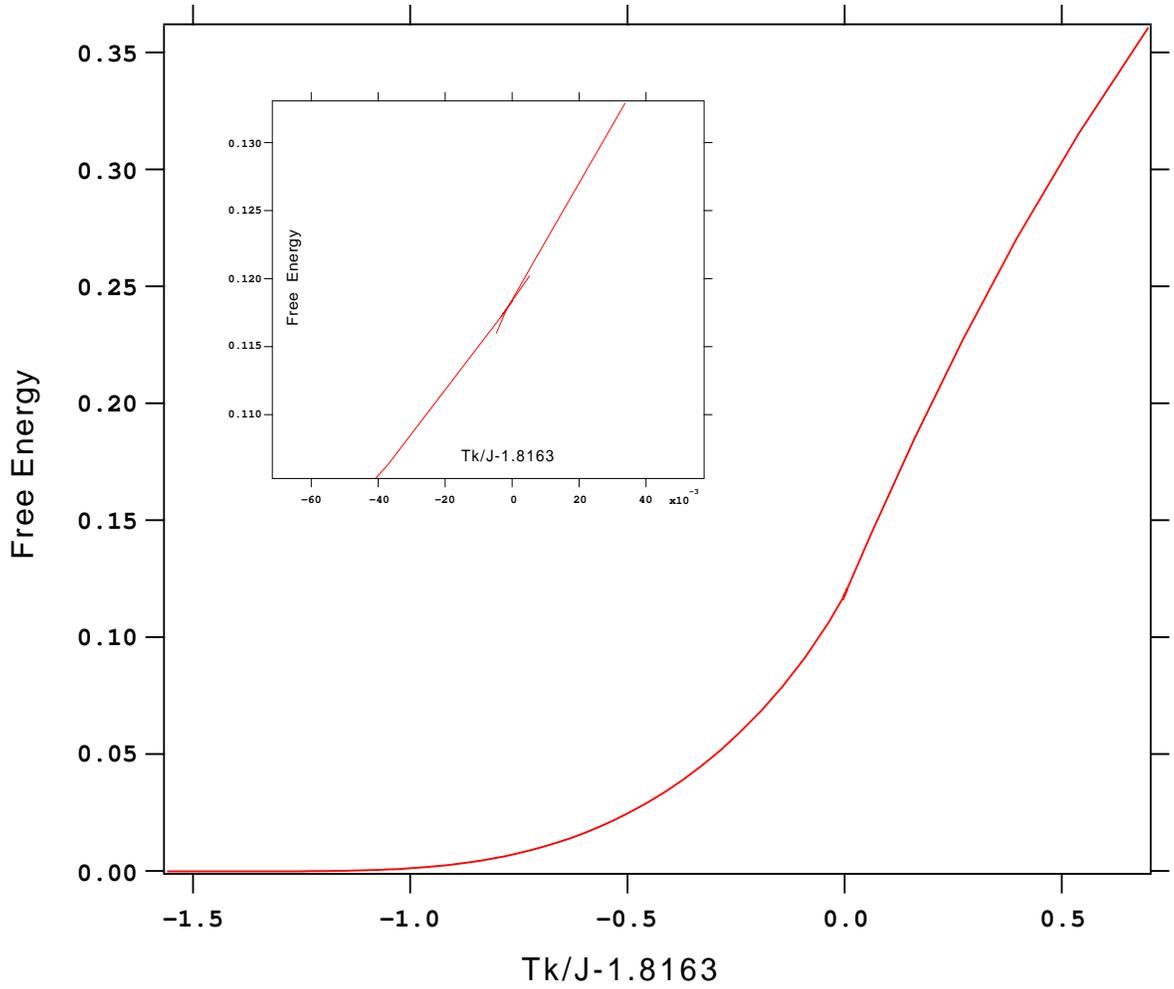

.